\documentclass{aastex}
\usepackage{spr-astr-addons}
\usepackage{url}

\RequirePackage{color}

\begin{document}

\title{{\sc profit}: a new alternative for emission-line {\sc pro}file {\sc fit}ting}
\slugcomment{Not to appear in Nonlearned J., 45.}
\shorttitle{{\sc profit}: a line-profile fitting routine}
\shortauthors{Rogemar A. Riffel}

\author{Rogemar A. Riffel}
\affil{Universidade Federal de Santa Maria, Departamento de F\'\i sica, Centro de Ci\^encias Naturais e Exatas, 
97105-900, Santa Maria, RS, Brazil}


\begin{abstract}
I briefly describe a simple routine for emission-line profiles fitting by Gaussian curves or Gauss-Hermite series. 
The {\sc profit} (line-{\sc pro}file {\sc fit}ting) routine represent a new alternative for use in fits data cubes, as the ones 
from Integral Field Spectroscopy or Fabry-Perot Interferometry, and may be useful  
 to better study the emission-line flux distributions and gas  kinematics 
in distinct astrophysical objects, such as the central regions of galaxies and star forming regions. 
The {\sc profit} routine is written in IDL language and is available at http://www.ufsm.br/rogemar/software.html.
  
The {\sc profit} routine was used to fit the [Fe\,{\sc ii}]$\lambda=1.257\,\mu$m emission-line profiles  
 for about 1800 spectra of the inner 350~pc of the Seyfert galaxy Mrk\,1066 obtained with Gemini NIFS and shows that the line
 profiles are better reproduced by Gauss-Hermite series than by
 the commonly used Gaussian curves. The two-dimensional map of the $h_3$ Gauss-Hermite moment shows its highest absolute values in regions 
close to the edge of the radio structure. These high values may be originated in an biconical outflowing gas associated with the radio jet -- 
 previously observed in the optical [O\, {\sc iii}] emission. The analysis of this kinematic component indicates that the radio jet leaves the 
center of the galaxy with the north-west side slightly oriented towards us and the south-east side away from us, being partially hidden
 by the disc of the galaxy. 
 
\end{abstract}



\section{Introduction} \label{intro}

Integral Field Spectroscopy (IFS) and Fabry-Perot Interferometry are powerful tools  to do a two-dimensional  
analysis of the physical properties of several types of astronomical objects, such as the central region 
of normal \citep[e.g.][]{rodrigues09,peletier07,emsellem07,diaz06} and active galaxies 
\citep[e.g.][]{davies09,davies07,hicks09,riffel10a,riffel09a,riffel09b,riffel08,riffel06,sb09a,sb09b,sb07,fathi06}, star forming 
regions \citep[e.g.][]{blum09,barbosa08} and young stellar objects \citep[e.g.][]{beck08,mcgregor07,takami07}. The final result of 
the data reduction of the above techniques is a data cube containing hundreds to thousands individual spectra to be analyzed, 
thus a common problem among the studies cited above is how to extract the information from these data cubes. A manually inspection 
of each spectrum is an exhaustive task and demand much time, so automated methods are needed to properly
 measure physical parameters from the data cubes.

The most common method used to study the gaseous distribution and kinematics is based on the fitting of the
emission-line profiles by Gaussian curves. Nevertheless, it is commonly reported in the literature 
the presence of asymmetries in the emission-line profiles, which are not well represented by Gaussian 
curves \citep[e.g.][]{barbosa09,komossa08,riffel09a,riffel10b}. Several methods have been developed to fit line profiles
 \citep[e.g][and {\sc iraf fitprofs} and {\sc splot} tasks]{sarzi06}, but most of these methods are based on the fitting of 
Gaussian (or Lorentzian) functions, in which the information of the wings of the line profile can be lost.
 A recent developed method to extract information from data cubes is the PCA tomography,  
which uses Principal Component Analysis (PCA) to transform the system of correlated coordinates into a system of uncorrelated coordinates 
ordered by principal components of decreasing variance  \citep{steiner09}. Nevertheless, in some cases the `traditional' line-profile 
fitting method must be additionally used to properly extract the flux distribution and kinematics of the emitting gas.  

In this work I present an automated line-{\sc pro}file {\sc fit}ting routine ({\sc profit}) to be used to extract the gaseous kinematics and 
flux distribution from fits data cubes. This routine is written in IDL\footnote{http://www.ittvis.com/} language and allows the fit of the observed profiles 
by  Gauss-Hermite series or  Gaussian curves. The fit of Gauss-Hermite series has been chosen because it preserve the velocity information of 
the emitting gas by the fitting of the wing of the emission-line profiles. Such information
 could be lost in the fit of a single Gaussian curve  for an asymmetric emission-line profile.  
Another advantage of the Gauss-Hermite profile is that it can be easily be implemented in an automated routine than multiple Gaussian fit -- 
 which would also preserve the  velocity information.

The paper is organized as follows. In Section~2 I present the formalism of the 
 Gauss-Hermite and Gaussian functions; Sec.~3 presents the {\sc profit} routine and in Sec.~4 I discuss an application of the routine for the 
Seyfert galaxy Mrk\,1066. Sec.~5 presents the final remarks of the present work.

\section{Gauss-Hermite \textit{versus} Gaussian}

The Gauss-Hermite series can be written as \citep[e.g.][]{vandermarel93,gerhard93,cappellari04}:
 \begin{equation}
 GH = \frac{A\alpha(w)}{\sigma}\sum_{j=0}^n h_jH_j(w)
 \label{gh}
 \end{equation}
 where
\begin{equation}
 w\equiv\frac{\lambda-\lambda_c}{\sigma}
\end{equation}
 and
\begin{equation}
 \alpha(w)=\frac{1}{\sqrt{2\pi}}e^{-w^2/2},
\end{equation}
 $A$ is the amplitude of the Gauss-Hermite 
 series, $\lambda_c$ is the peak wavelength, $h_j$ are the Gauss-Hermite moments and $H_j(w)$ are the Hermite polynomials.

\begin{figure}[t]
 \centering
 \begin{minipage}{1.0\linewidth}
 \includegraphics[scale=0.27]{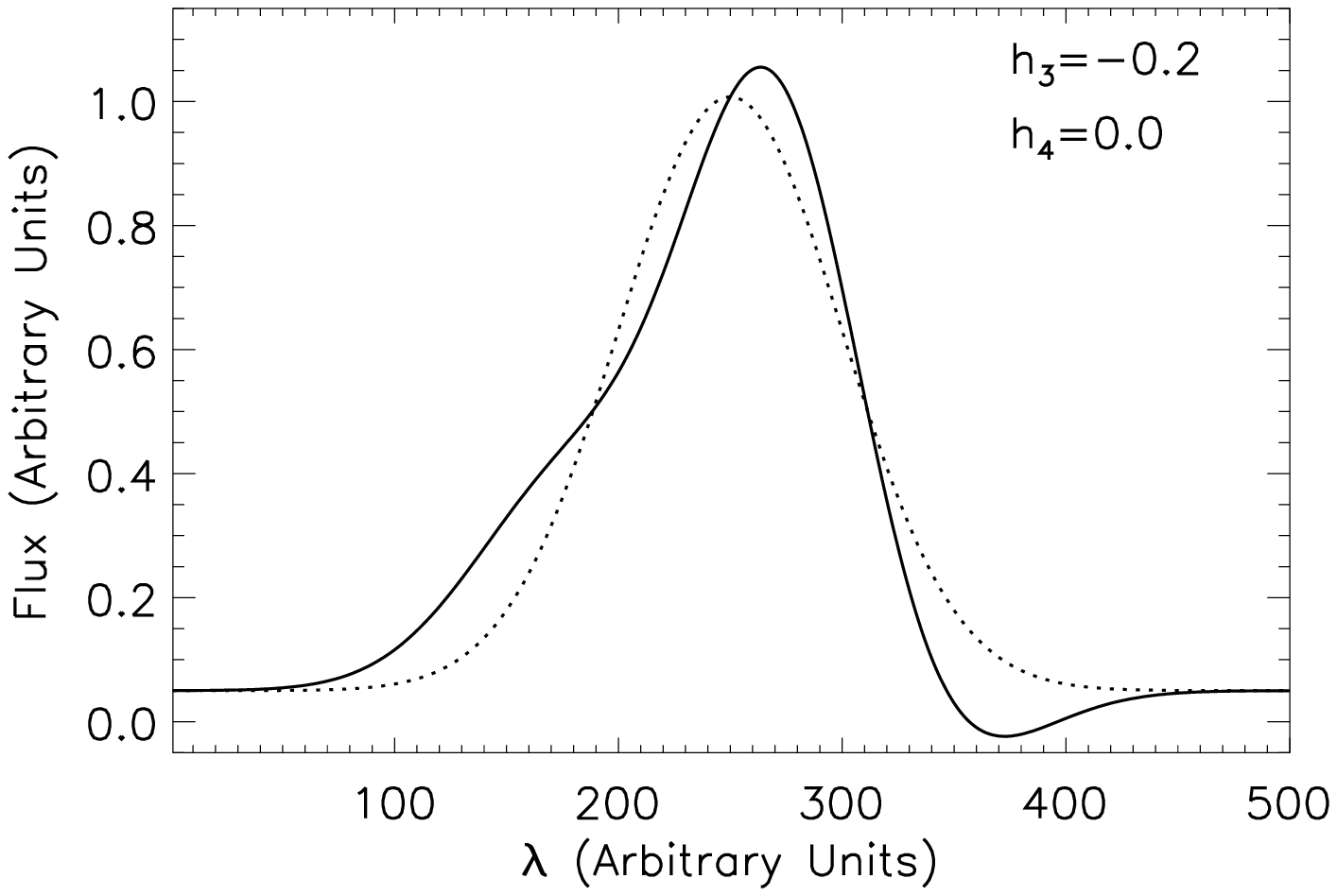}
 \includegraphics[scale=0.27]{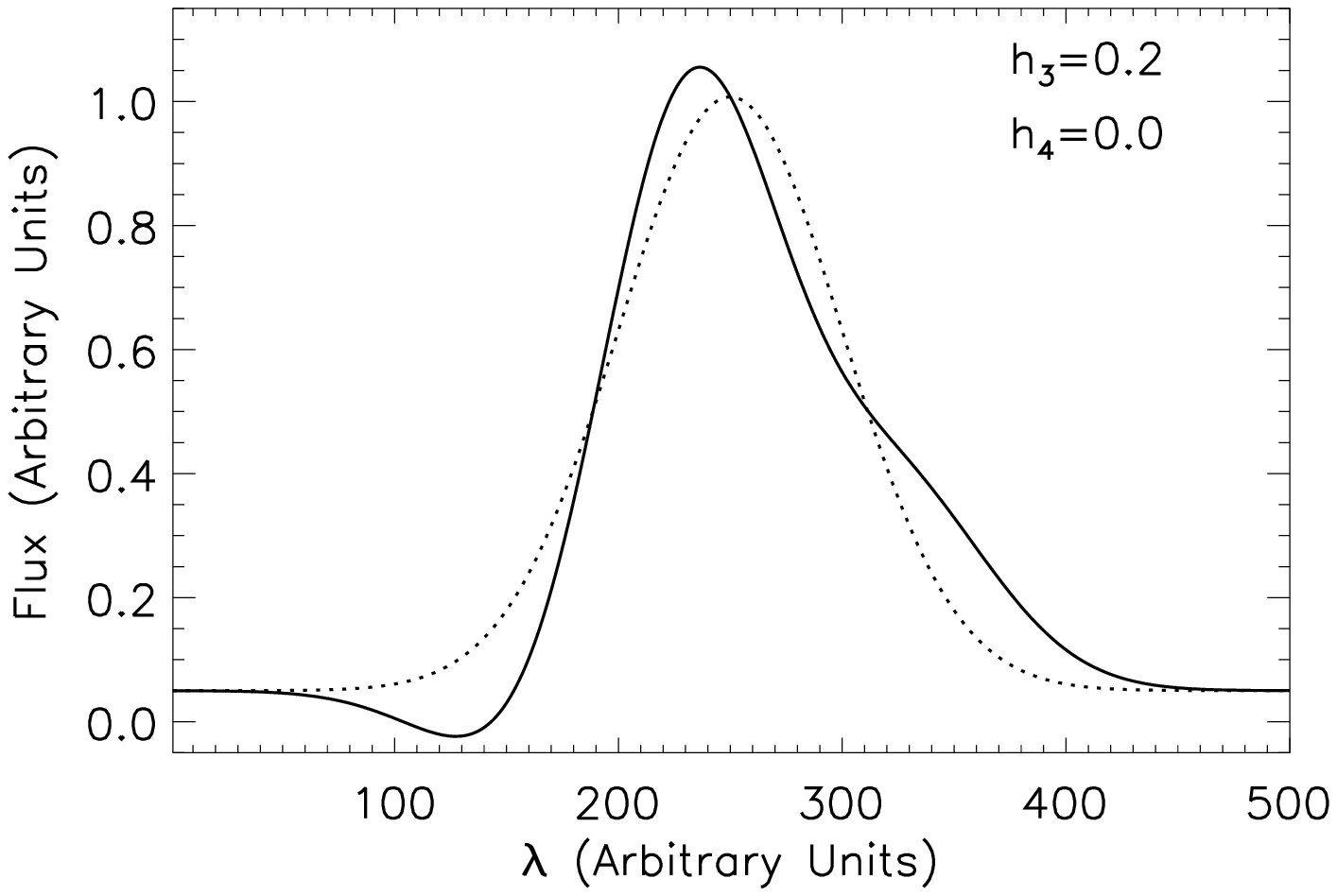}
 \end{minipage}
 \begin{minipage}{1\linewidth}
 \includegraphics[scale=0.27]{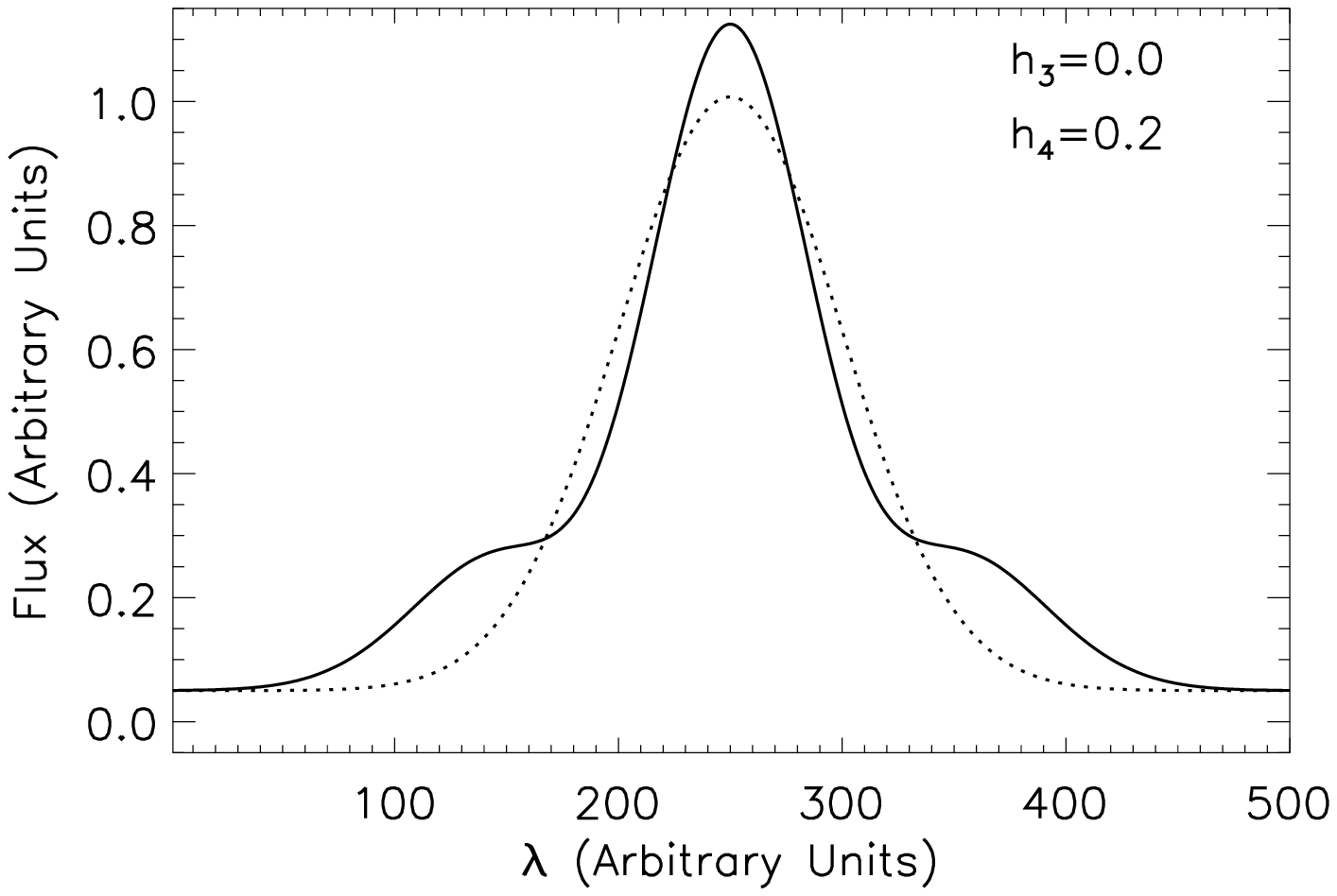}
 \includegraphics[scale=0.27]{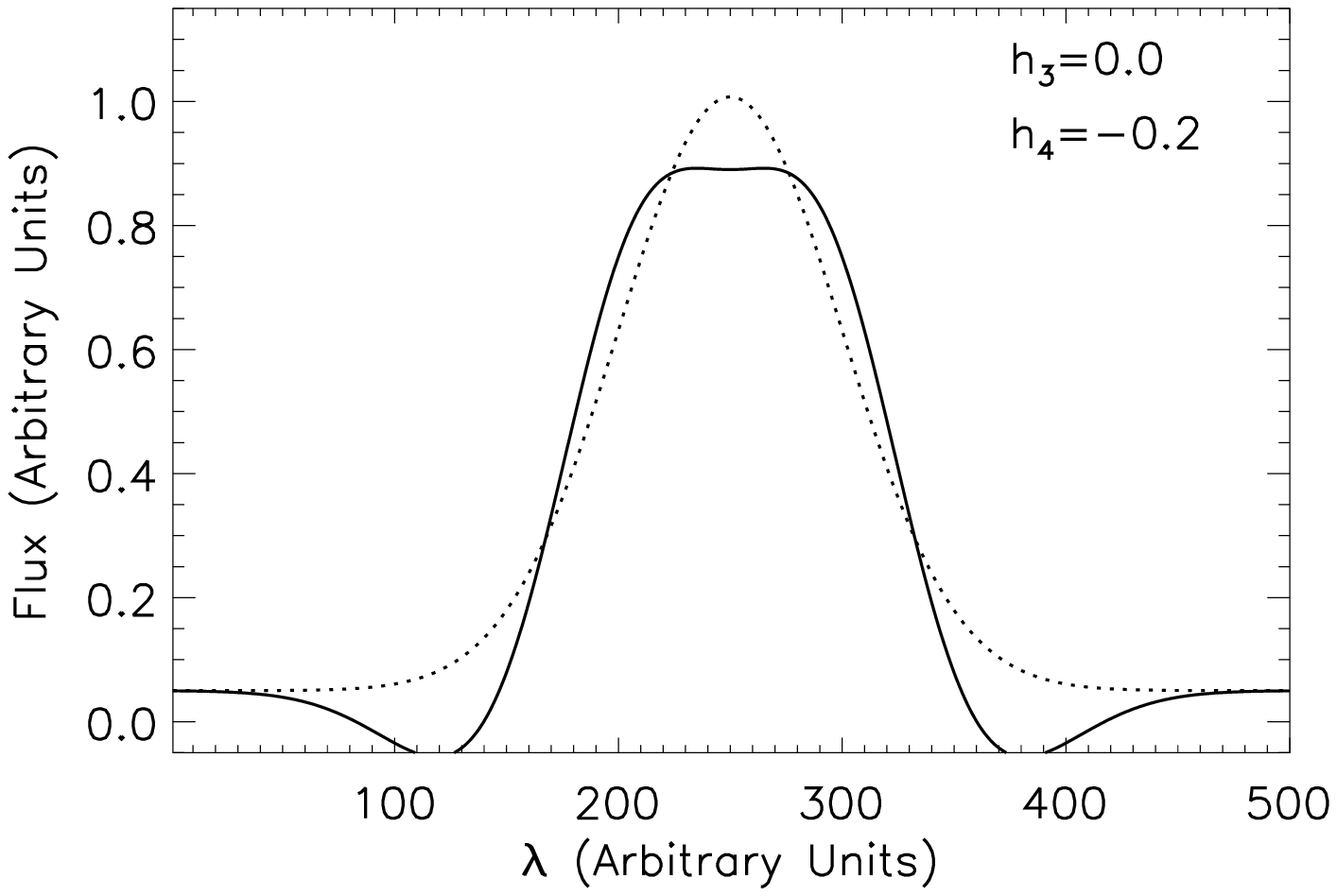}
 \end{minipage}
 \begin{minipage}{1\linewidth}
 \includegraphics[scale=0.27]{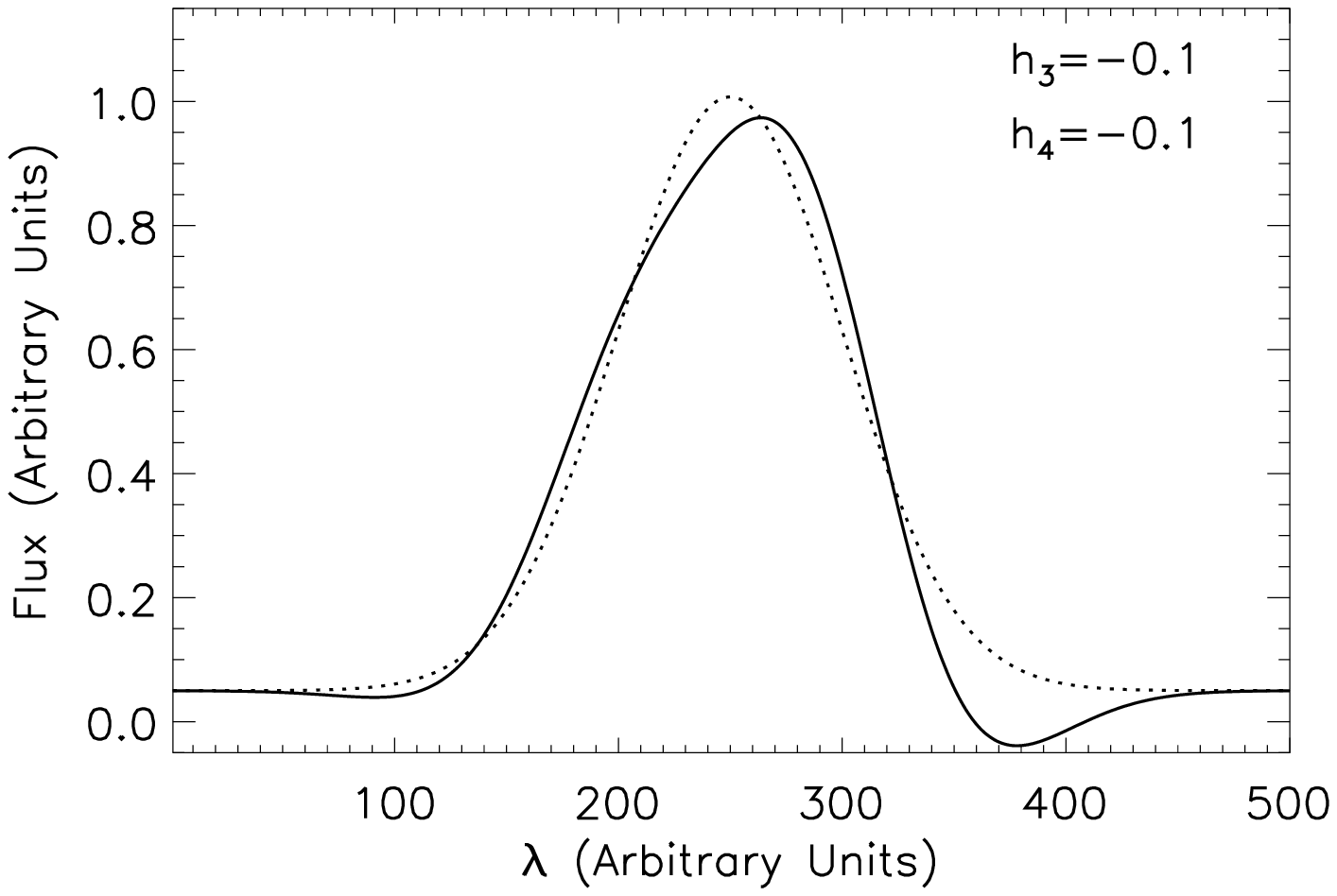}
 \includegraphics[scale=0.27]{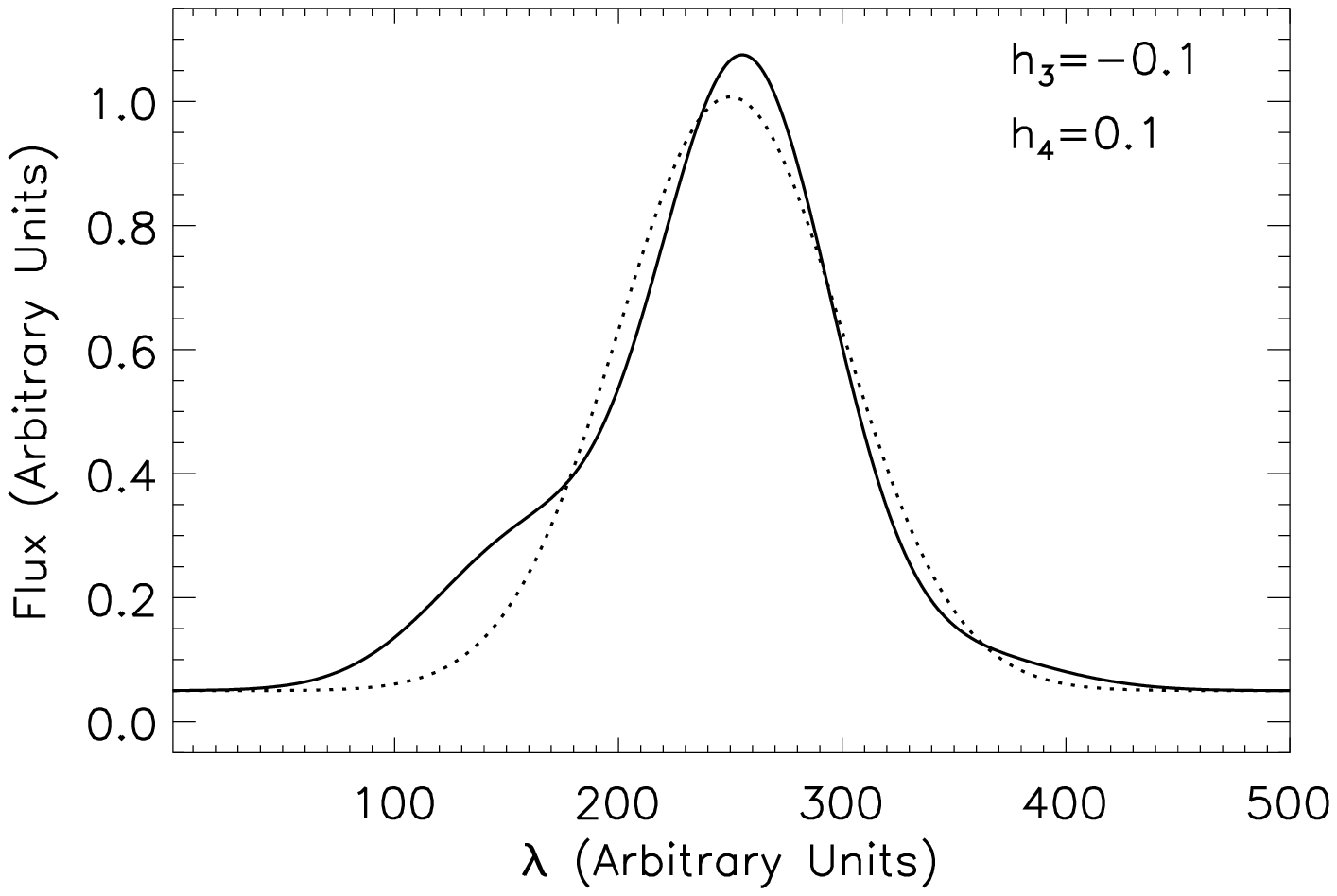}
 \end{minipage}
 \caption{Comparisson of Gaussian curves (dotted lines)
 with Gauss-Hermite series (continuous lines) for the $h_3$ 
and $h_4$ values shown at the top-left corner of each panel. The amplitude, central 
wavelength and $\sigma$ are the same for both functions and 
have arbitrary values.} 
 \label{sample}  
 \end{figure}

If the emission-line profile is similar to a Gaussian  we can truncate the sum on $n=4$ and assume  $h_0=H_0(w)=1$, $h_1=h_2=0$ 
\citep{vandermarel93}. This is a good approximation if the emission-line presents an asymmetric profile, such as the blue or red wings 
 frequently observed for the line emission from ionized gas in the narrow line region of active galaxies \citep{riffel10a,riffel09a,komossa08}. 
 Using the approximation above, the  Eq.\,\ref{gh} can be written as

 \begin{equation}
  GH=\frac{A\alpha(w)}{\sigma}\left[1 + h_3 H_3(w) + h_4 H_4(w)\right],
 \label{ghfit}
 \end{equation}
 where 
 \begin{equation}
  H_3(w)=\frac{1}{\sqrt{6}}(2\sqrt{2}w^3-3\sqrt{2}w) 
\end{equation}
and 

 \begin{equation} 
 H_4(w)=\frac{1}{\sqrt{24}}(4w^4-12w^2+3).
 \end{equation}

The $h_3$ Gauss-Hermite moment measures asymmetric deviation from a Gaussian profile, 
such as blue or red wings, while the $h_4$ moment quantify the peakiness of the profile, 
with $h_4 > 0$ for a more peaked and $h_4 < 0$ for a broader profile than a Gaussian curve. 
A particular case of Eq.~\ref{ghfit} is $h_3=h_4=0$, when it becomes a Gaussian curve. 
In Figure~\ref{sample}, I present a sample of profiles for Gauss-Hermite series 
with distinct $h_3$ and  $h_4$ values (continuous lines) and Gaussian curves (dotted lines). 
The amplitude, central wavelength and $\sigma$ are the same for Gaussian and Gauss-Hermite functions 
and have arbitrary values. The $h_3$ and  $h_4$ moments are shown at the top-left corner of each panel and have values 
typically observed for the narrow line region of active galaxies \citep[e.g.][]{riffel10b}.
 The comparisson of observed emission-line profiles,  
such as those of Mrk\,1066 from \citet{riffel10a} (see Sec.~\ref{application}), with 
the profiles shown in Figure~\ref{sample}  suggests that the observations are better reproduced by Gauss-Hermite series than by Gaussian curves
 in most cases.

\section{The {\sc profit} routine}

The model of each emission-line profile is constructed by the sum of Eq.~\ref{ghfit} with a linear equation, in order to 
represent the underlying continuum emission. The resulting equation contains seven free parameters 
($A,\, \lambda_c, \sigma,\,h_3\,h_4$ plus two parameters for the linear equation), which can 
 be determined by fitting the line profiles. In case of Gaussian fitting the $h_3$ and $h_4$ are fixed at zero 
and thus the remaining 5 parameters may be obtained from the fit of the observed profile. These 
parameters can be obtained by solving a Least-squares problem.

The {\sc profit} routine was written in IDL language and performs the fit of the observed profile using  
 the {\sc mpfit}\footnote{The MPFIT routine can be obtained from the Markwardt IDL Library at http://cow.physics.wisc.edu/~craigm/idl/idl.html}
 routine, which the is the MINPACK1 implementation \citep{more80}
of the Levenberg-Marquardt method for nonlinear least-squares problems. The IDL language was chosen because it is 
extensively used in astronomy and allows read the 
data cube from standard fits format to an array using the NASA--Goddard Space Flight Center IDL Astronomy User's 
Library\footnote{The IDL Astronomy User's Library can be obtained from http://idlastro.gsfc.nasa.gov}. The algorithm recovers the emission-line 
flux distribution and kinematics as follows: 

\noindent 1- Initial guesses for the centroid wavelength and velocity dispersion are given by the user. The initial guesses for 
$h_3$ and $h_4$ are fixed at zero;

\noindent 2- The input data cube in  standard fits format (in which the spatial dimensions are in the x and y-axis and spectral 
pixels are in the z-axis) is converted into an array.

\noindent Next steps are done individually for each spectrum:

\noindent 3- Calculates the spectral region to be fitted using the initial guess for the centroid wavelength and  the spectral information
(spectral sampling and initial wavelength) contained in the 
header of the data cube fits file. 

\noindent 4- Normalizes the spectrum by its maximum value and obtain initial guesses for the parameters of the linear equation and amplitude 
of the Gauss-Hermite series (or Gaussian);

\noindent 5- Performs the nonlinear least-squares fitting of the observed profile by the adopted the model using the  Levenberg-Marquardt method;

\noindent 6- Writes the solution to the output file;

\noindent 7- If $\chi^2$ is less than a maximum value (defined by the user) the fitted parameters are used as initial guesses 
for the fit of the next spectrum. Otherwise uses the initial guesses of item 1;

\noindent 8- Repeats items 3--8 for all spectra.

\noindent 9- Writes the solutions to a Multiple Extensions FITS (MEF) file. The output MEF file will 
contain 7 extensions containing the: [0] emission-line flux distribution; [1] centroid velocity field; [2] velocity dispersion map;
 [3] $h_3$ map; [4] $h_4$ map; [5] reduced $\chi^2$ map defined as
\begin{equation}
 \chi^2=\sum_p\frac{ (O_p-M_p)^2}{\sigma_O^2}\frac{1}{(N-N_{\rm par})},
\end{equation}
where $O_p$ is the observed spectra, $M_p$ is the best fit model, $\sigma_O^2$ is the variance of the observed spectra, $N$ is the number 
of spectral pixels ($p$) used in the fit and  $N_{\rm par}$ is the number of free parameters; 
and [6] flux distribution obtained directly by integration the emission line profile 
and subtracting a continuum obtained by the average of continuum regions at both sides of the line profile.

\section{A First Application} \label{application}

\begin{figure}[t]
 \centering
 \includegraphics[scale=0.4]{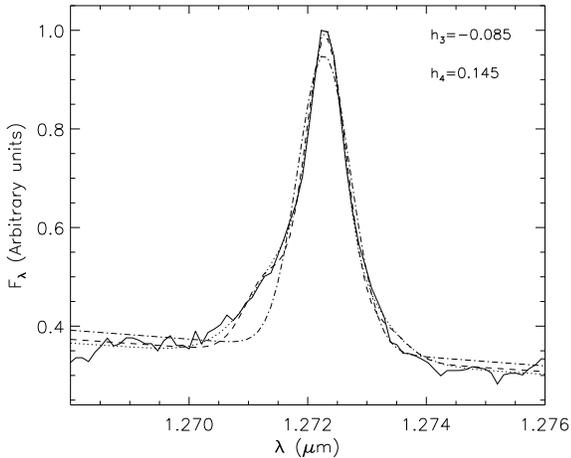}
 \caption{Example of fitting using {\sc profit} routine for the [Fe\,{\sc ii}]$\lambda1.2570\,\mu$m emission-line profile for a spectrum extracted 
within an aperture of 0.3$\times$0.3\,arcsec$^2$ at 1 arcsec north-west of the nucleus of Mrk\,1066. The continuous line represents 
the observed profile, the dashed line the resulting Gauss-Hermite fit, the dot-dashed line the fit of a single Gaussian and the 
dotted line the resulting fit by two Gaussian curves.} 
 \label{fe}  
 \end{figure}

\begin{figure}[t]
 \centering
 \includegraphics[scale=0.4]{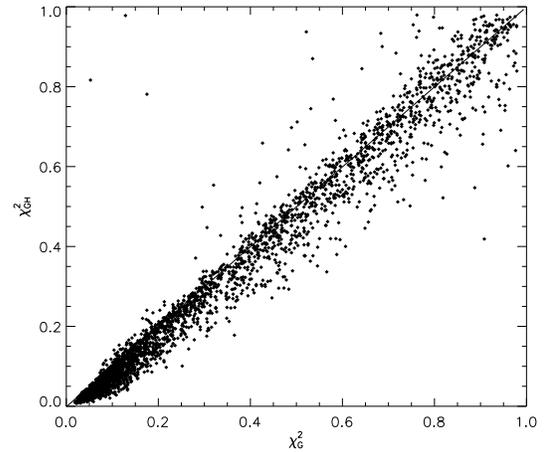}
 \caption{Comparisson of the resulting $\chi^2$ obtained from the fitting of Gauss-Hermite series (y-axis) 
and Gaussian curves (x-axis) to the [Fe\,{\sc ii}]\,$\lambda$1.2570\,$\mu$m emission-line profile of $\sim$1800 spectra of
 the inner 700$\times$700\,pc$^2$ of Mrk\,1066.} 
 \label{chi}  
 \end{figure}

In Fig.~\ref{fe} I present an example of the use of {\sc profit} to fit the emission-line profile of 
[Fe\,{\sc ii}] at $\lambda=1.257\,\mu$m  at  1$^{\prime\prime}$
 north-west of the nucleus of the Seyfert galaxy Mrk\,1066.  
This spectrum was extracted from 
 Gemini's Near-infrared Integral Field Spectrograph (NIFS) observations (program ID: GN-2008A-Q-30) 
within an aperture of 0.3$\times$0.3\,arcsec$^2$ (see \citet{riffel10a} for 
a description of the observations and data reduction procedures). The observed profile is shown 
as a continuous line,the fitting of Gauss-Hermite series as a dashed line, the fitting of a single Gaussian as 
a dot-dashed line and the two Gaussian curves fit as a dotted line. 
 The best Gauss-Hermite fit is obtained for a central wavelength $\lambda_c=12722.6\pm0.2\,\AA$, a velocity dispersion of
$\sigma=114.3\pm8.8$\,km\,s$^{-2}$ and higher order Gauss-Hermite moments  $h_3=-0.085\pm0.017$ and $h_4=0.145\pm0.014$
 and resulted in a $\chi^2=0.014$.  
The best fit obtained using a Gaussian curve has
$\chi^2=0.049$ is obtained for $\lambda_c=12722.8\pm0.2\,\AA$ and $\sigma=106.2\pm11.9$\,km\,s$^{-2}$. 
Although the values obtained for $\lambda_c$ and $\sigma$ from Gauss-Hermite and Gaussian functions are similar, Fig.~\ref{fe} clearly shows 
that the observed profile is better reproduced by Gauss-Hermite series than by a Gaussian curve -- the blue wing present 
on the observed profile is not reproduced by the Gaussian curve. This conclusion is also evidenced 
by the lower $\chi^2$ value obtained for the Gauss-Hermite fitting. 
In the other hand, the resulting $\chi^2$ obtained for the fitting of the [Fe\,{\sc ii}] emission-line profile by 
two Gaussian curves  $\chi^2_{2G} = 0.011$ is similar to those obtained from the 
Gauss-Hermite fitting. However, in an automated fitting routine it is hard to decide when an emission-line profile 
should be fitted by a single Gaussian and when it should be fitted by multiple Gaussian curves \citep[e.g.][]{sb09b}, while for 
 Gauss-Hermite series this decision is done simple by varying the $h_3$ and $h_4$ moments.

\begin{figure}
 \centering

 \includegraphics[scale=0.5]{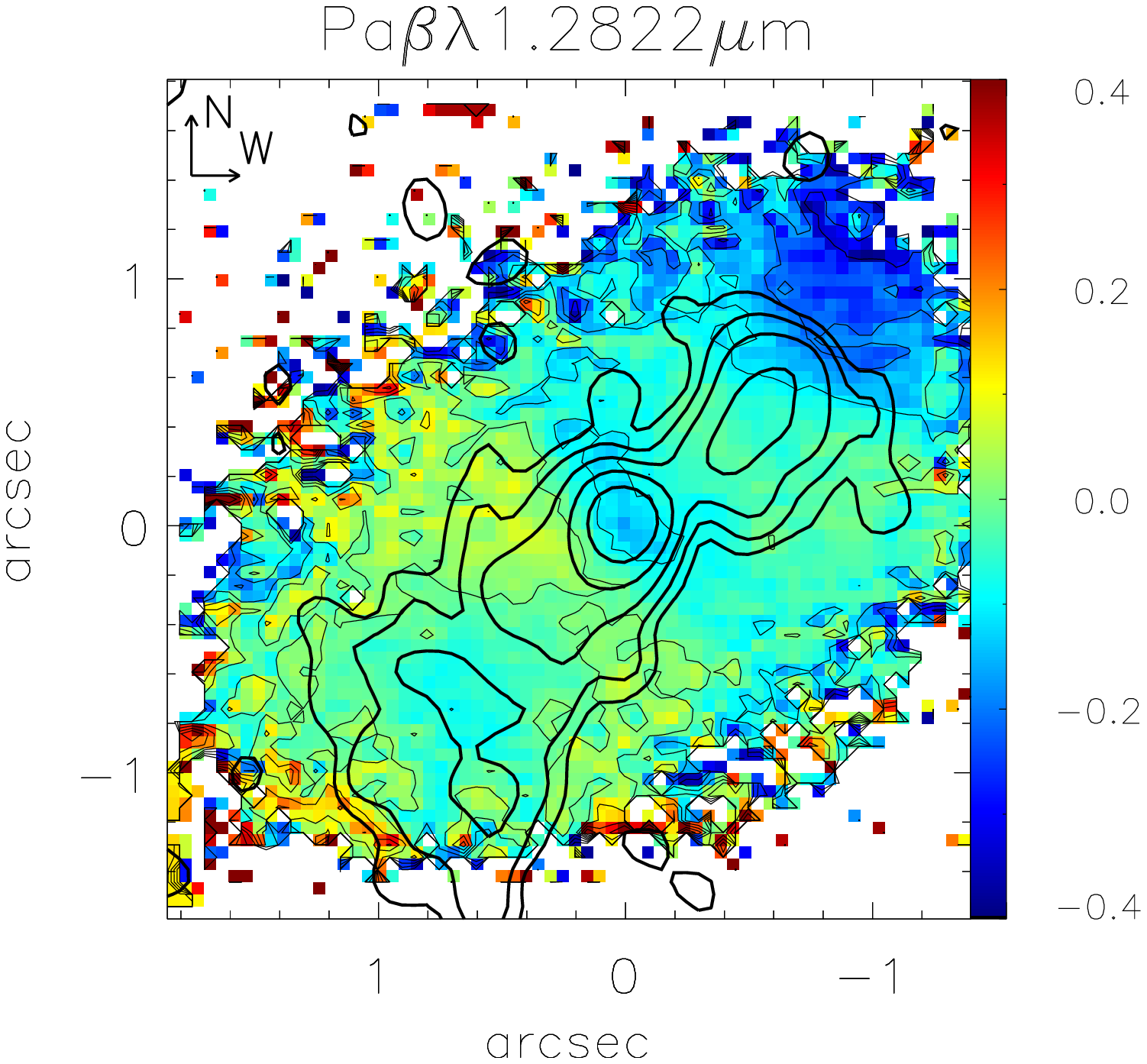}
 \includegraphics[scale=0.5]{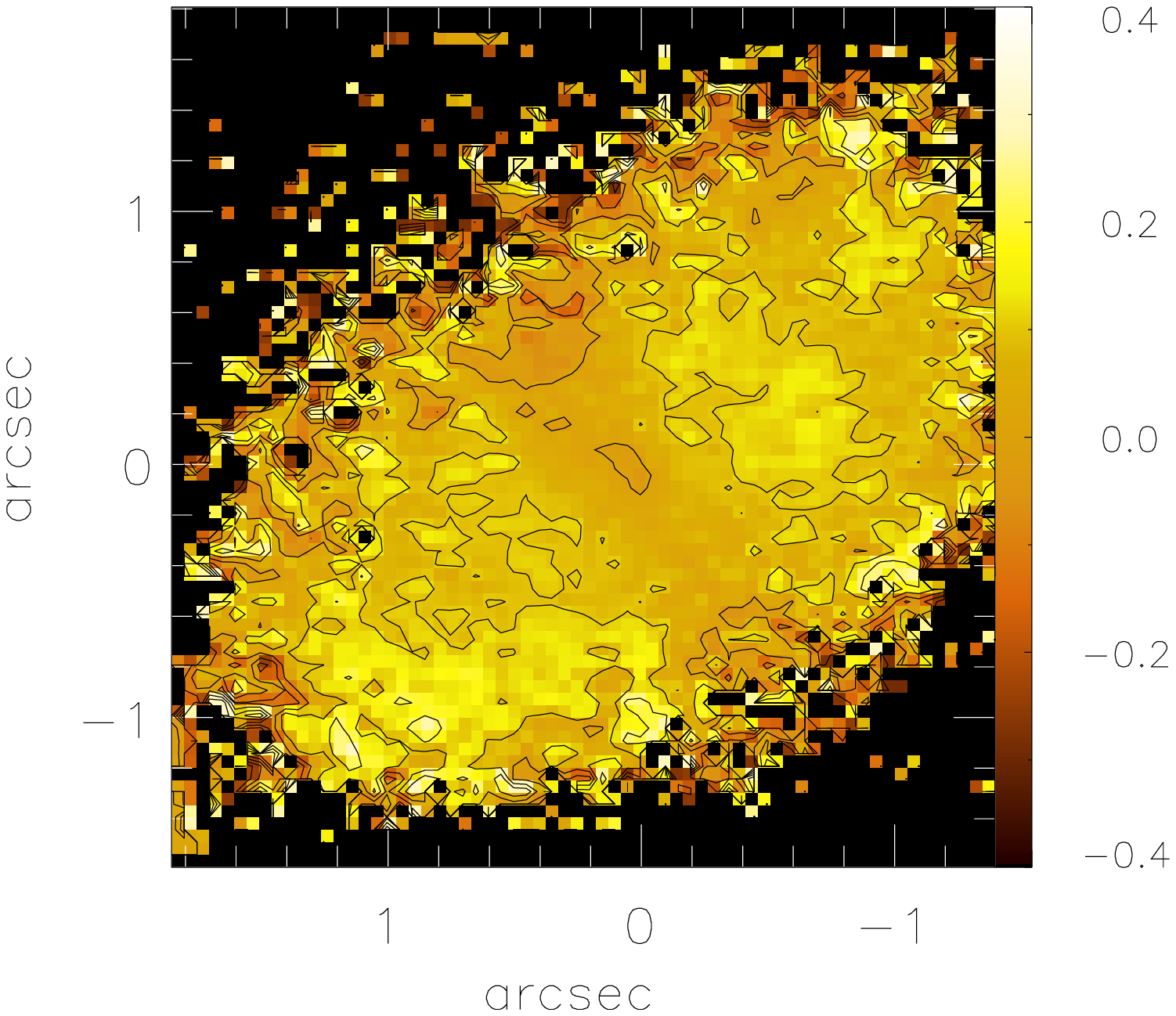}
 \caption{Top: $h_3$ Gauss-Hermite moment map for Pa$\beta$ emission line on the inner 700$\times$700\,pc$^2$ of Mrk\,1066. The thick black contours 
are from the radio continuum emission of \citet{nagar99}. Bottom: $h_4$ Gauss-Hermite moment map.} 
 \label{h3}  
 \end{figure}

In \citet{riffel10b} we used {\sc profit} to study the gaseous kinematics of the inner 700$\times$700\,pc$^2$ of Mrk\,1066 
using Gemini NIFS observations. The profiles of [P\,{\sc ii}]\,$\lambda$1.1886\,$\mu$m, 
 [Fe\,{\sc ii}]\,$\lambda$1.2570\,$\mu$m, Pa$\beta$ and H$_2\,\lambda$2.1218$\mu$m emission lines were fitted 
by Gauss-Hermite series in order to obtain the centroid velocity field, velocity dispersion map and $h_3$ and $h_4$ maps 
for each emission line. 
The same observations presented in \citet{riffel10b} were used to fit the [Fe\,{\sc ii}]\,$\lambda$1.2570\,$\mu$m by Gaussian curves in 
order to compare the resulting fit with the ones obtained using Gauss-Hermite series. Figure~\ref{chi} presents the comparisson 
of the $\chi^2$ values from the Gauss-Hermite fitting ($\chi^2_{GH}$ -- y-axis) with the ones obtained for the fitting of 
Gaussian curves ($\chi^2_G$ -- x-axis) for $\sim$1800 spectra of the inner 700$\times$700\,pc$^2$ of Mrk\,1066 extracted within
 apertures of 0.05$\times$0.05\,arcsec$^2$. As 
observed in this figure, $\chi^2_{GH}$ is smaller than $\chi^2_G$ for most spectra, indicating that the [Fe\,{\sc ii}] line profile 
in the central region of Mrk\,1066 is better reproduced by Gauss-Hermite series than by Gaussian curves. A similar behavior is observed 
for [P\,{\sc ii}], H$_2$ and H emission lines at the same spatial region.

In order to illustrate the importance of properly map the emission line profile wings Fig.~\ref{h3} presents a map for the
 $h_3$ Gauss-Hermite moment obtained for the  Pa$\beta$ emission in the central region of Mrk\,1066. This map is simmilar to the ones 
obtained for the [Fe\,{\sc ii}] and [P\,{\sc ii}] emission lines \citep{riffel10b}, showing several regions 
with values different than zero, indicating that the Pa$\beta$ emission-line profile presents asymmetric deviation from a Gaussian 
curve, such as blue (negative values) and red (positive values) wings. In case of Gaussian fitting this information could be lost!
The smallest values of  up to $-$0.3
are observed at $\approx1^{\prime\prime}$ north-west of the nucleus in a regions close to the edge of the radio structure (thick black contours). 
Some high values are also observed near to the edge of the radio jet to south-east of the nucleus. 
The presence of wings have also been observed by \citet{knop01} for the near-IR emission lines using long slit spectroscopy along the  PA=135$^\circ$.
 These wings may be originated in an outflowing gas component driven by the radio jet,   
with the north-west side slightly oriented towards us and the south-east side away from us, being partially hidden by the disc of the galaxy. 
This interpretation is supported by the near-IR emission line kinematic maps, which show that the near-IR emitting gas presents at least three 
kinematics components: a rotating disk, an inflowing gas component and an outflowing component \citep{riffel10b}. The above interpretation is also in good agreement with optical observations of the 
[O\,{\sc iii}] emission, which seems to being originated in a bi-cone oriented along the same position angle PA$\approx$135$^\circ$ -- approximately the 
same orientation of the radio jet \citep{bower95}. 

The $h_4$ map, shown in the bottom panel of Fig.~\ref{h3},  presents values near to zero in most locations of the central region of Mrk\,1066. However, 
some positive $h_4$-values are observed co-spatially with regions of lower velocity dispersion, indicating that part of the Pa$\beta$ emission originates from 
a colder gas than those which produces most of the line emission. For more details on the gaseous kinematics 
of the inner 350~pc radius of Mrk\,1066 see \citet{riffel10b}.

\section{Final Remarks}

I presented a new routine to fit emission-line profiles from fits data cubes  
using Gauss-Hermite series or Gaussian curves, which provides a new alternative to the study of the emission-line flux distribution and kinematics for several 
astronomical objects. The {\sc profit} routine is written in IDL 
and is available at http://www.ufsm.br/rogemar/software.html. The main advantages 
of {\sc profit} compared with previous methods are:

\begin{itemize}

\item  It allows the fitting of the line profiles by  
 Gauss-Hermite series as an alternative to the `traditional' Gaussian curves used in most studies and thus better describe the kinematics of the 
emitting gas. 
\item It is automated and can be applied directly on the final data cubes fits files 
 from observations using most integral field units and Fabry-Perot interferometers.

\end{itemize}

A first application has been discussed for the case of near-IR emission-line profiles from the inner 350~pc of the Seyfert galaxy Mrk\,1066. The main 
scientific conclusions are:
\begin{itemize}
\item The near-IR emission line profiles for this galaxy are better reproduced by Gauss-Hermite series than by Gaussian curves, as indicated by the 
smaller $\chi^2$ obtained for the former.

\item The two-dimensional map for the $h_3$ Gauss-Hermite shows that the near-IR emission lines present asymmetric profiles,
 which can be explained as being originated in an outflowing gas driven by the radio jet. 

\end{itemize}

\section*{Acknowledgments}
I thank the referee for valuable suggestions which helped to improve the present paper. 
Based on observations obtained at the Gemini Observatory, which is operated by the
Association of Universities for Research in Astronomy, Inc., under a cooperative agreement
with the NSF on behalf of the Gemini partnership: the National Science Foundation (United
States), the Science and Technology Facilities Council (United Kingdom), the
National Research Council (Canada), CONICYT (Chile), the Australian Research Council
(Australia), Minist\'erio da Ci\^encia e Tecnologia (Brazil) 
and Ministerio de Ciencia, Tecnología e Innovaci\'on Productiva  (Argentina).

\end{document}